\documentclass[letterpaper,12pt,notoc]{JHEP3}
\def\pd{\partial}
\def\Slash{{\!\!\!\!/}}

\usepackage{graphicx}
\usepackage{amsmath}
\usepackage{amssymb}

\preprint{ \hbox{}\hfill arXiv:1006.4997}

\title{3D gauged supergravity from SU(2) reduction of $N=1$
6D supergravity }

\author{Edi Gava$^a$, Parinya Karndumri$^{a,\, b}$ and K. S. Narain$^c$\\
$^a$INFN, Sezione di Trieste, Italy\\
$^b$International School for Advanced Studies (SISSA), via Bonomea
265, 34136 Trieste, Italy \\
$^c$The Abdus Salam International Centre for Theoretical Physics,
Strada Costiera 11, 34100 Trieste, Italy \\
E-mail: \email{gava$@$ictp.it}, \email{karndumr$@$sissa.it},
\email{narain$@$ictp.it}}

\abstract{We obtain Yang-Mills $SU(2)\times G$ gauged supergravity
in three dimensions from $SU(2)$ group manifold reduction of (1,0)
six dimensional supergravity coupled to an anti-symmetric tensor
multiplet and gauge vector multiplets in the adjoint of $G$. The
reduced theory is consistently truncated to $N=4$ 3D supergravity
coupled to $4(1+\textrm{dim}\, G)$ bosonic and $4(1+\textrm{dim}\,
G)$ fermionic propagating degrees of freedom. This is in contrast to
the reduction in which there are also massive vector fields. The
scalar manifold is $\mathbf{R}\times \frac{SO(3,\, \textrm{dim}\,
G)}{SO(3)\times SO(\textrm{dim}\, G)}$, and there is a $SU(2)\times
G$ gauge group. We then construct $N=4$ Chern-Simons $(SO(3)\ltimes
\mathbf{R}^3)\times (G\ltimes \mathbf{R}^{\textrm{dim}G})$ three
dimensional gauged supergravity with scalar manifold
$\frac{SO(4,\,1+\textrm{dim}G)}{SO(4)\times SO(1+\textrm{dim}G)}$
and  explicitly show that this theory is on-shell equivalent to the
Yang-Mills $SO(3)\times G$ gauged supergravity theory obtained from
the $SU(2)$ reduction, after integrating out the scalars and gauge
fields corresponding to the translational symmetries
$\mathbf{R}^3\times \mathbf{R}^{\textrm{dim}\, G}$.}

\begin{document}
\section{Introduction}
Three dimensional Chern-Simons gauged supergravities have a very
rich structure and admit various types of gauging including
non-semisimple and complex gauge groups \cite{nicolai1, nicolai2,
nicolai3, dewit}. It has been shown in \cite{dewit, csym} that
non-semisimple Chern-Simons gaugings with gauge group
$G\ltimes\mathbf{R}^{\textrm{dim}\, G}$  are on-shell equivalent to
semisimple Yang-Mills gaugings with gauge group $G$, including
Chern-Simons couplings. This result makes it possible to obtain some
of the Chern-Simons gauged supergravities with non-semisimple gauge
groups from dimensional reductions of higher dimensional theories.
For example, in \cite{PopeSU2,PopeSU22} the dimensional reduction of
pure (1,0) six-dimensional supergravity on an $SU(2)$ group manifold
\cite{SS} has been shown to give rise to a three dimensional gauged
supergravity with $SU(2)$ Yang-Mills gauge group plus $SU(2)$
massive Chern-Simons vector fields and scalars in $GL(3,{\bf
R})/SO(3)$, whose action has the structure essentially found in
\cite{csym}. Also, the $N=8$ theories studied in \cite{ns2} and
\cite{gkn, HS1}, with $SO(4)$ and $SO(4)^2$ Yang-Mills gaugings
respectively, are expected to arise from six dimensional (2,0)
theory on $S^3$ and from IIB theory on $S^3\times S^3\times S^1$,
respectively. Notice that in these examples, on the 3D supergravity
side one can allow different gauge couplings for the two $SU(2)$'s
in $SO(4)$, whereas from the $S^3$ dimensional reduction there is a
single gauge coupling for the gauge fields arising from the
isometries of $S^3$.
\\ \indent
In this paper, we will consider an $N=4$, 3D gauged supergravity
where the scalar manifold is a single quaternionic space
$\frac{SO(4,1+\textrm{dim}\, G)}{SO(4)\times SO(1+\textrm{dim}\,
G)}$ and $(SO(3)\ltimes \textbf{R}^3)\times
(G\ltimes\mathbf{R}^{\textrm{dim}\, G})$ Chern-Simons gauging, where
$G$ is an arbitrary semisimple group. We will show that this theory
can be obtained from an $SU(2)$ reduction of a (1,0) six-dimensional
supergravity coupled to a tensor multiplet and Yang-Mills multiplets
of the gauge group $G$.
\\ \indent
The six dimensional (1,0) gauged supergravity has been constructed
in \cite{nishino1} and extended to couple to $n_T$ anti-symmetric
tensor multiplets, $n_V$ vector multiplets and $n_H$ hypermultiplets
in \cite{nishino}. The theory has been completed with quartic
fermion terms in \cite{Riccioni}. For earlier constructions of six
dimensional (1,0) supergravity, we refer the reader to \cite{Romans,
Sagnotti}. We only consider the truncation of this extended theory
to the ungauged theory, with $n_T=1$ and $n_H=0$, coupled to $G$
Yang-Mills gauge fields. After reducing to three dimensions, we will
show that the resulting theory is equivalent to the Chern-Simons
gauged theory by reversing the procedure of \cite{csym}.
\\ \indent
The paper is organized as follows. In section \ref{6Dsugra}, we
review (1,0) six dimensional supergravity in order to set up our
notations. In section \ref{reduction}, we will perform the $SU(2)$
group manifold reduction of (1,0) six dimensional supergravity
coupled to an anti-symmetric tensor and $G$ Yang-Mills multiplets
and obtain $SU(2)\times G$ Yang-Mills gauged supergravity in three
dimensions. The resulting theory contains $4(1+\textrm{dim}\, G)$
bosons and $4(1+\textrm{dim}\, G)$ fermions with the scalar manifold
being $\textbf{R}\times \frac{SO(3,\, \textrm{dim}\, G)}{SO(3)\times
SO(\textrm{dim}\, G)}$. While it is known that the most general
$SU(2)$ reduction, including massive vector fields, is consistent,
the novel feature of our work is that we make a further truncation
by removing the massive vector fields and show that it is
consistent. In section \ref{CSYM}, we construct an $N=4$
Chern-Simons $(SO(3)\ltimes \mathbf{R}^3)\times (G\times
\mathbf{R}^{\textrm{dim}G})$ gauged theory with a scalar manifold
$\frac{SO(4,\, 1+\textrm{dim}G)}{SO(4)\times SO(1+\textrm{dim}G)}$
and show that it is indeed equivalent to $SO(3)\times G$ Yang-Mills
gauged theory with scalar manifold $\mathbf{R}\times \frac{SO(3,\,
\textrm{dim}G)}{SO(3)\times SO(\textrm{dim}G)}$ after removing
$3+\textrm{dim}G$ scalars corresponding to the translational
symmetries. We finally give some conclusions and comments in section
\ref{conclusions}.
\section{$N=(1,0)$ six dimensional supergravity}\label{6Dsugra}
In this section, we briefly review and set up our notations for
(1,0) six dimensional supergravity coupled to an antisymmetric
tensor and $G$ Yang-Mills multiplets. Pure (1,0) supergravity does
not have a covariant action because of the self-duality of the three
form field strength. Due to the anti-self duality of the 3-form
field strength in the tensor multiplet, the coupled theory does
admit a Lagrangian formulation. Six dimensional supergravity coupled
to $n_T$ anti-self dual tensor multiplets, $n_V$ Yang-Mills vector
multiplets and $n_H$ hypermultiplets has been constructed in
\cite{nishino}. We are interested in the case of $n_T=1$ which
admits a supersymmetric action. The theory considered here contains
$N=1$ supergravity multiplet, one antisymmetric tensor multiplet and
$\textrm{dim}\, G$ Yang-Mills multiplets of an arbitrary gauge group
$G$. We also assume that the group $G$ commutes with the $SU(2)\sim
Sp(1)$ R-symmetry group. The field content in this case is given by
the graviton $e^{\hat{M}}_M$, gravitino $\psi^A_M$, third rank
anti-symmetric tensor ${G_3}_{MNP}$, scalar $\theta$, spin
$\frac{1}{2}$ fermion $\chi$, $G$ gauge fields $A_M^I$ with $I=1, 2,
\ldots , \textrm{dim}\, G$ and the $G$ gauginos $\lambda^I$. The six
dimensional spacetime indices are $M, N=0,\ldots 5$ with the tangent
space indices $\hat{M}, \hat{N}=0,\ldots 5$ while $A, B=1, 2$ are
$Sp(1)$ R-symmetry indices. The Lagrangian for this theory, with
$\tilde{v}^z=0$, is given by \cite{nishino}
\begin{eqnarray}
e^{-1}\mathcal{L}&=&\frac{1}{4}R-\frac{1}{12}e^{2\theta}{G_3}_{MNP}{G_3}^{MNP}-\frac{1}{4}\partial_M
\theta \partial^M \theta -\frac{1}{2}\bar{\psi}_M\Gamma^{MNP}D_N
\psi_P\nonumber \\
& &-\frac{1}{2}\bar{\chi}\Gamma^MD_M\chi-\frac{1}{4}v^ze^\theta
F^I_{MN}F^{IMN}-v^ze^\theta\bar{\lambda}^I\Gamma^MD_M\lambda^I\nonumber \\
& &+\frac{1}{2}v^ze^\theta\bar{\chi}\Gamma^{MN} \lambda^IF^I_{MN}
+\frac{1}{2}\bar{\psi}_M\Gamma^N\Gamma^M\chi\partial_N\theta-\frac{1}{2}v^ze^\theta\bar{\psi}_M\Gamma^{NP}\Gamma^M\lambda^IF^I_{NP}
\nonumber \\
& &-\frac{1}{24}e^\theta
{G_3}_{MNP}\big[\bar{\psi}^L\Gamma_{[L}\Gamma^{MNP}\Gamma_{Q]}\psi^Q-2\bar{\psi}_L\Gamma^{MNP}\Gamma^L\chi
-\bar{\chi}\Gamma^{MNP}\chi\nonumber \\
& &+2v^ze^\theta \bar{\lambda}^I\Gamma^{MNP}\lambda^I \big]
\end{eqnarray}
where $e=\sqrt{-g}$. We use the same metric signature as in
\cite{nishino}, $(-+++++)$. The supersymmetry transformations for
various fields are \cite{nishino}
\begin{eqnarray}
\delta e^{\hat{M}}_M&=&\bar{\epsilon}\Gamma^{\hat{M}}\psi_M, \nonumber\\
\delta \psi_M &=& D_M \epsilon +\frac{1}{24}e^\theta
\Gamma^{NPQ}\Gamma_M {G_3}_{NPQ}\epsilon
-\frac{1}{16}\Gamma_M\chi\bar{\epsilon}\chi-\frac{3}{16}\Gamma^N\chi\bar{\epsilon}\Gamma_{MN}\chi\nonumber
\\ & & +\frac{1}{32}\Gamma_{MNP}\chi
\bar{\epsilon}\Gamma^{NP}\chi-\frac{1}{16}v^ze^\theta(18\lambda^I\bar{\epsilon}\Gamma_M\lambda^I
-2\Gamma_{MN}\lambda^I\bar{\epsilon}\Gamma^N\lambda^I\nonumber
\\ & &+\Gamma_{NP}\lambda
\bar{\epsilon} \Gamma_M^{\phantom{a}NP}\lambda),\nonumber \\
\delta b_{MN}&=&2v^zA^I_{[M}\delta
A^I_{N]}-e^{-\theta}\bar{\epsilon}
\Gamma_{[M}\psi_{N]}+\frac{1}{2}e^{-\theta}\bar{\epsilon}\Gamma_{MN}\chi,\nonumber
\\
\delta \theta &=&\bar{\epsilon}\chi,\nonumber\\
\delta \chi&=&\frac{1}{2}\Gamma^M\partial_M\theta\epsilon
-\frac{1}{12}e^\theta\Gamma^{MNP}{G_3}_{MNP}\epsilon
+\frac{1}{2}v^ze^\theta
\Gamma^M\lambda^I(\bar{\epsilon}\Gamma_M\lambda^I),\nonumber\\
\delta A^I_M&=&-\bar{\epsilon}\Gamma_M\lambda^I,\nonumber \\
\delta
\lambda^I_A&=&\frac{1}{4}\Gamma^{MN}F^I_{MN}\epsilon_A-C_z^{-1}v^ze^\theta
\bar{\chi}_{(A}\lambda_{B)} \epsilon^B.\label{6Dsusyvar}
\end{eqnarray}
Notice that by our assumption on the gauge group $G$ on the r.h.s.
of $\delta \lambda^I_A$, the term $C^{AB}\epsilon_B$ is missing. One
can see, using the formalism developed in the next section, that the
presence of this term would imply a reduction of supersymmetry (if
any) of the three dimensional theory. The bosonic field equations
are given by \cite{nishino}
\begin{eqnarray}
R_{MN}-\frac{1}{2}g_{MN}R-\frac{1}{3}e^{2\theta}\big(3{G_3}_{MPQ}{G_3}_N^{\phantom{a}PQ}-\frac{1}{2}g_{MN}{G_3}_{PQR}{G_3}^{PQR}\big)\nonumber
\\-\partial_M\theta
\partial^M\theta+\frac{1}{2}g_{MN}\partial_P\theta\partial^P\theta-e^\theta
\big(2F^{IP}_MF^I_{NP}-\frac{1}{2}g_{MN}F^I_{PQ}F^{IPQ}\big)&=&0, \label{Eins}\\
e^{-1}\partial_M(eg^{MN}\partial_N\theta)-\frac{1}{2}e^\theta
F^I_{MN}F^{IMN}-\frac{1}{3}e^{2\theta}{G_3}_{MNP}{G_3}^{MNP}&=&0,
\label{thetaeq}\\
D_N(ee^\theta
F^{IMN})+ee^{2\theta}G^{MNP}F^I_{NP}&=&0,\label{Feq}\\
D_M(ee^{2\theta}{G_3}^{MNP})&=&0\label{Geq}.
\end{eqnarray}
We also choose $v^z=1$ from now on. The three form field strength is
\begin{equation}
G_3=db+F^I\wedge A^I-\frac{1}{6}g_2f_{IJK}A^I\wedge A^J\wedge A^K
\end{equation}
where $g_2$ and $f_{IJK}$ are coupling and structure constants of
the gauge group $G$, respectively. The equations of motion for
various fermions can be found in \cite{nishino}. We will not repeat
them here because they will not be needed in this work. In the next
section, we will give the reduction ansatz and perform the
dimensional reduction of this theory on the $SU(2)$ group manifold.
\section{3D $SU(2)\times G$ gauged supergravity from (1,0) six dimensional supergravity on
$S^3$}\label{reduction} In this section, we study Kaluza-Klein
reduction of the (1,0) six dimensional supergravity coupled to an
anti-symmetric and Yang-Mills vector multiplets with a gauge group
$G$ on $SU(2)$ group manifold. The result is the $N=4$, $SU(2)\times
G$ gauged supergravity in three dimensions. The (1,0) six
dimensional supergravity has been obtained from various
compactifications of string and M theory e.g. \cite{sen},
\cite{green}. These compactifications have been used to study many
aspects of string dualities in six dimensions e.g. \cite{witten},
\cite{duff}.
\subsection{Reduction ansatz on $SU(2)$ group manifold}
We now give our reduction ansatz. We will put a hat on all the six
dimensional fields from now on. We use the following reduction
ansatz:
\begin{eqnarray}
d\hat{s}^2&=&e^{2f}ds^2+e^{2g}h_{\alpha \beta}\nu^\alpha \nu^\beta
\, , \nonumber \\ \hat{A}^I&=&A^I+A^I_\alpha \nu^\alpha, \qquad
\nu^\alpha=\sigma^\alpha-g_1A^\alpha \, ,\nonumber \\
\hat{F}^I&=&d\hat{A}^I+\frac{1}{2}g_2f_{IJK}\hat{A}^J\wedge\hat{A}^K\nonumber
\\ &=& F^I-g_1A^I_\alpha F^\alpha +\mathcal{D}A^I_\alpha\wedge
\nu^\alpha + \frac{1}{2}(g_2A^J_{\alpha} A^K_{\beta}
f_{IJK}-\epsilon_{\alpha \beta \gamma}A^I_\gamma)\nu^\alpha \wedge
\nu^\beta
\end{eqnarray}
where the $SU(2)\times G$ covariant derivative is given by
\begin{equation}
\mathcal{D}A^I_\alpha=dA^I_\alpha +g_1\epsilon_{\alpha
\beta\gamma}A^\beta A^I_\gamma+g_2f_{IJK}A^JA^K_\alpha .
\end{equation}
The three dimensional field strength
$F^I=dA^I+\frac{1}{2}g_2f_{IJK}A^J\wedge A^K$. From the metric, we
can read off the vielbein components
\begin{equation}
\hat{e}^a=e^fe^a,\, \, \hat{e}^i=e^gL^i_\alpha \nu^\alpha \, \,
\textrm{with} \, \, h_{\alpha\beta}=L^i_\alpha L^i_\beta.
\end{equation}
The left-invariant $SU(2)$ 1-forms $\sigma^\alpha$ satisfy
\begin{equation}
d\sigma^\alpha=-\frac{1}{2}\epsilon_{\alpha\beta\gamma}\sigma^\beta\wedge\sigma^\gamma\,
.
\end{equation}
The $\epsilon_{\alpha\beta\gamma}$ and $f_{IJK}$ are the $SU(2)$ and
$G$ structure constants, respectively. The metric $h_{\alpha\beta}$
and a $(3\times 3)$ matrix $L^i_\alpha$ are unimodular. The spin
connections are given by \cite{PopeSU2}
\begin{eqnarray}
\hat{\omega}_{ab}&=&\omega_{ab}+e^{-f}(\partial_bf\eta_{ac}-\partial_af\eta_{bc})\hat{e}^c+\frac{1}{2}g_1e^{g-2f}F^i_{ab}\hat{e}^i,\nonumber\\
\hat{\omega}_{ai}&=&-e^{-f}P_{aij}\hat{e}^j-e^{-f}\partial_ag\hat{e}^i+e^{g-2f}F^i_{ab}\hat{e}^b,\nonumber\\
\hat{\omega}_{ij}&=&e^{-f}Q_{aij}\hat{e}^a+\frac{1}{2}e^{-g}(T^{kl}\epsilon_{ijl}+T^{jl}\epsilon_{ikl}-T^{il}\epsilon_{jkl})\hat{e}^k\label{spinconnection}
\end{eqnarray}
where
\begin{eqnarray}
P_{aij}&=&\frac{1}{2}\big[(L^{-1})^\alpha_iD_aL^j_\alpha+(L^{-1})^\alpha_jD_aL^i_\alpha\big]=\frac{1}{2}(L^{-1})^\alpha_i(L^{-1})^\beta_j
D_ah_{\alpha\beta},\nonumber\\
Q_{aij}&=&\frac{1}{2}\big[(L^{-1})^\alpha_iD_aL^j_\alpha-(L^{-1})^\alpha_jD_aL^i_\alpha\big],\nonumber\\
F^i&=&L^i_\alpha F^\alpha, \qquad T^{ij}=L^i_\alpha L^j_\alpha,
\qquad DL^i_\alpha=dL^i_\alpha
-g_1\epsilon_{\alpha\beta\gamma}A^\gamma L^i_\beta.
\end{eqnarray}
We use the same conventions as in \cite{PopeSU2} namely
\begin{eqnarray}
F^\alpha &=&
dA^\alpha+\frac{1}{2}g_1\epsilon_{\alpha\beta\gamma}A^\beta\wedge
A^\gamma, \nonumber \\
DF^\alpha&=&dF^\alpha+g_1\epsilon_{\alpha\beta\gamma}A^\beta\wedge
A^\gamma=0, \nonumber \\
D\nu^\alpha&=&d\nu^\alpha+g_1\epsilon_{\alpha\beta\gamma}A^\beta\wedge
\nu^\gamma=-g_1F^\alpha-\frac{1}{2}\epsilon_{\alpha\beta\gamma}\nu^\beta\wedge
\nu^\gamma.
\end{eqnarray}
The indices $(M,\hat{M})$ reduce to $(\mu,a)$ in three dimensions
while the $S^3$ part is described by indices $(\alpha,i)$. The
ansatz for $\hat{G}_3$ is
\begin{eqnarray}
\hat{G}_3&=&h\varepsilon_3+a\epsilon_{\alpha\beta\gamma}\nu^\alpha\wedge\nu^\beta\wedge\nu^\gamma+
\epsilon_{\alpha\beta\gamma}C^\alpha\wedge\nu^\beta\wedge\nu^\gamma+H^\alpha\wedge\nu^\alpha\nonumber
\\ & &+\hat{F}^I\wedge \hat{A}^I-\frac{1}{6}g_2f_{IJK}\hat{A}^I\wedge \hat{A}^J\wedge
\hat{A}^K.\label{Gansatz}
\end{eqnarray}
The first line in \eqref{Gansatz} is the $d\hat{b}$ which must be
closed. This requires that
\begin{equation}
H^\alpha=2DB^\alpha-6ag_1F^\alpha.
\end{equation}
We also choose the one form $C_\alpha=\frac{1}{2}A^I_\alpha A^I$ to
further simplify the ansatz and truncate the vector field $C^\alpha$
out. Putting all together, we end up with the following $\hat{G}_3$
ansatz
\begin{eqnarray}
\hat{G}_3&=&\tilde{h}\varepsilon_3+\bar{F}^\alpha\wedge \nu^\alpha
+\frac{1}{2}K_{\alpha \beta}\wedge\nu^\alpha \wedge\nu^\beta
\nonumber \\ & &
+\frac{1}{6}\tilde{a}\epsilon_{\alpha\beta\gamma}\nu^\alpha\wedge\nu^\beta\wedge\nu^\gamma
\end{eqnarray}
where $\tilde{h}=h\varepsilon_3+\tilde{F}^I\wedge
A^I-\frac{1}{6}g_2A^I\wedge A^J\wedge A^Kf_{IJK}$. We have defined
the following quantities
\begin{eqnarray}
\bar{F}^\alpha &=&A^I_\alpha (\tilde{F}^I+F^I)-6ag_1 F^\alpha,
\qquad \tilde{F}^I=F^I-g_1A^I_\alpha F^\alpha, \nonumber \\
K_{\alpha
\beta}&=&A^I_\beta\mathcal{D}A^I_\alpha-A^I_\alpha\mathcal{D}A^I_\beta,
\nonumber \\
\tilde{a}&=&6a-A^I_\alpha A^I_\alpha+\frac{1}{3}g_2A^3, \qquad
A^3\equiv A^I_\alpha A^J_\beta A^K_\gamma
f_{IJK}\epsilon_{\alpha\beta\gamma},
\end{eqnarray}
and $a$ is a constant. The ansatz for the Yang-Mills fields can be
rewritten as
\begin{equation}
\hat{F}^I=\tilde{F}^I+\mathcal{D}A^I_\alpha\wedge
\nu^\alpha+\frac{1}{2}\mathcal{F}^I_{\alpha\beta}\nu^\alpha\wedge\nu^\beta
\end{equation}
where $\mathcal{F}^I_{\alpha\beta}=g_2A^J_\alpha A^K_\beta
f_{IJK}-A^I_\gamma \epsilon_{\alpha\beta\gamma}$. \\ \indent The
volume form in three dimensions is defined by
\begin{equation}
\varepsilon_3=\frac{1}{6}e^{3f}\epsilon_{abc}e^a\wedge e^b\wedge
e^c\equiv e^{3f}\omega_3\, .
\end{equation}
The six dimensional gamma matrices decompose as \cite{PopeSU2}
\begin{eqnarray}
\Gamma^{\hat{A}}&=&(\Gamma^a,\Gamma^i), \qquad \Gamma^a=\gamma^a
\otimes
\mathbb{I}_2 \otimes \sigma_1, \nonumber \\
\Gamma^i&=& \mathbb{I}_2 \otimes \gamma^i\otimes \sigma_2, \qquad
\Gamma_7=\mathbb{I}_2 \otimes \mathbb{I}_2 \otimes \sigma_3
\nonumber \\
\gamma^{abc}&=&\epsilon^{abc},\qquad \gamma^{ijk}=i\epsilon^{ijk},
\qquad \{\gamma_a,\gamma_b\}=2\eta_{ab},\, \,\,
\{\gamma_i,\gamma_j\}=2\delta_{ij}.
\end{eqnarray}
The conventions are $\eta_{AB}=(-+++++)$, $\eta_{ab}=(-++)$ and
$\epsilon^{012}=\epsilon^{345}=1$. We further choose
\begin{equation}
\gamma^0=i\tilde{\sigma}^2, \,\, \gamma^1=\tilde{\sigma}^1,\, \,
\gamma^2=\tilde{\sigma}^3, \qquad \gamma^{i}=\tau^i
\end{equation}
where $\tilde{\sigma}^i$, $\tau^i$, $i=1,2,3$ are the usual Pauli
matrices. Also the chirality condition
$\Gamma_7\epsilon^A=\epsilon^A$ becomes
$\mathbb{I}_2\otimes\mathbb{I}_2\otimes\sigma_3\epsilon^A=\epsilon^A$.
\\ \indent
Before proceeding further, let us count the number of degrees of
freedom. Table \ref{table1} shows all three dimensional fields
arising from the six dimensional ones.
\TABLE{\begin{tabular}{|c|c|c|}
  \hline
  6D fields & 3D fields & 3D number of degrees of freedom  \\
  \hline
  $\hat{g}_{MN}$ & $g_{\mu\nu}$  & non propagating \\
    & $A_\mu ^\alpha$  & $3$ \\
    & $h_{\alpha\beta}$ & 5 \\
    & $g$ & 1
  \\ \hline
  $\hat{b}_{MN}$ & $b_{\mu\nu}$ & non propagating\\
   & $b_{\mu\alpha}$ & 3 \\
   & $b_{\alpha\beta}$ & 3
  \\ \hline
  $\hat{\theta}$ & $\theta$ & 1 \\ \hline
  $\hat{A}^I_M$ & $A^I_\mu$ & $ \textrm{dim} G$\\
   & $A^I_\alpha$ &  $3\,\textrm{dim} G$\\ \hline
  $\hat{\psi}_M$ & $\psi_\mu$ & non propagating \\
  & $\psi _i$ & 12\\ \hline
  $\hat{\lambda}^I$ & $\lambda^I$ & $4\, \textrm{dim} G$ \\ \hline
  $\hat{\chi}$ & $\chi$ & 4 \\ \hline
 \end{tabular}\caption{Three dimensional fields and the associated number of degrees of
freedom.}}\label{table1} From table \ref{table1}, there are
$16+4\textrm{dim}G$ bosonic and $16+4\textrm{dim}G$ fermionic
degrees of freedom in the full reduced theory. In this counting,
each six dimensional fermion gives rise to 4 three dimensional
fermions. In the reduction of the six dimensional theory, the
component $\hat{b}_{\mu\alpha}$ will give rise to massive vector
fields in three dimensions. Our goal is to truncate this theory to
obtain a three dimensional $N=4$ gauged supergravity involving only
gravity, scalars and gauge fields without massive vector fields. The
resulting theory will have $4(1+\textrm{dim}\, G)$ bosonic and
$4(1+\textrm{dim}\, G)$ fermionic propagating degrees of freedom. To
achieve this, we need to truncate 12 degrees of freedom out. From
the $\hat{G}_3$ ansatz expressed entirely in terms of gauge fields,
scalars coming from the gauge fields in six dimensions and
constants, we see that all the fields coming from $\hat{b}_{MN}$
have been truncated out. This accounts for 6 degrees of freedom. We
will see below that $h_{\alpha\beta}$ and $\theta$, comprising 6
degrees of freedom, will be truncated, too.
\\
\indent In the fermionic sector, we find that the truncation is
given by
\begin{equation}
\hat{\psi}_i-\frac{1}{2}\Gamma_i\hat{\chi}-2e^{\theta-g}A^I_\alpha
(L^{-1})^\alpha_i \hat{\lambda}^I=0.\label{fermiontruncation}
\end{equation}
Indeed, this removes 12 fermionic degrees of freedom. We have
checked that this truncation is compatible with supersymmetry to
leading order in fermions. We refer the readers to appendix
\ref{detail} for the detail of this computation. From appendix
\ref{detail}, we find that
\begin{equation}
\delta\hat{\psi}_i-\frac{1}{2}\Gamma_i\delta\hat{\chi}-2e^{\theta-g}A^I_\alpha
(L^{-1})^\alpha_i\delta\hat{\lambda}^I=0
\end{equation}
provided that
\begin{equation}
h_{\alpha\beta}=e^{\theta-2g}(12a\delta_{\alpha \beta}-2A^I_\alpha
A^I_\beta)\equiv e^{\theta-2g}N_{\alpha\beta}.
\label{bosonitruncation}
\end{equation}
This is the truncation in the bosonic sector. From
\eqref{bosonitruncation}, it follows that
\begin{equation}
\theta=2g-\frac{1}{3}\ln N \label{thetaTruncation}
\end{equation}
where $N\equiv \textrm{det}(N_{\alpha\beta})$. Also, from
\eqref{bosonitruncation}, it can be easily checked that
\begin{equation}
\delta [h_{\alpha\beta}-e^{\theta-2g}(12a\delta_{\alpha
\beta}-2A^I_\alpha A^I_\beta)]=0\label{deltah}
\end{equation}
to leading order in fermions by using \eqref{fermiontruncation}. The
detail can be found in appendix \ref{detail}. So, the relation
\eqref{bosonitruncation} is compatible with supersymmetry. Equations
\eqref{bosonitruncation} and \eqref{thetaTruncation} give another
truncation in the bosonic sector and remove 6 degrees of freedom.
The bosonic degree of freedoms are then given by $1+3\,
\textrm{dim(G)}$ scalars, $g$ and $A^I_\alpha$, and
$3+\textrm{dim(G)}$ vectors, $A^\alpha$ and $A^I$. So, the reduced
theory contains $4(1+\textrm{dim}G)$ propagating degrees of freedom
and involves only gravity, scalars and vector gauge fields.
\\ \indent We now check the consistency of the six dimensional field equations. It is
convenient to rewrite equations \eqref{thetaeq}, \eqref{Feq} and
\eqref{Geq} in differential forms. We find that these equations can
be written as
\begin{eqnarray}
\hat{\mathcal{D}}(e^{2\hat{\theta}}\hat{*}\hat{G}_3)&=&0,\label{G3eq}\\
\hat{\mathcal{D}}(e^{\hat{\theta}}
\hat{*}\hat{F}^I)-2e^{2\hat{\theta}}\hat{*}\hat{G}_3\wedge
\hat{F}^I&=&0,\label{FIeq} \\
\hat{d}\hat{*}\hat{d}\hat{\theta}+e^{\hat{\theta}}
\hat{*}\hat{F}^I\wedge
\hat{F}^I+2e^{2\hat{\theta}}\hat{*}\hat{G}_3\wedge\hat{G}_3&=&0\label{THeq}
\, .
\end{eqnarray}
In order to obtain the canonical Einstein-Hilbert term in three
dimensions, we choose $f=-3g$ from now on. Before giving equations
of motion, we give here the Hodge dual of $\hat{F}^I$ and
$\hat{G}_3$
\begin{eqnarray}
\hat{*}\hat{F}^I&=&\frac{1}{3!}e^{6g}*\tilde{F}^I\epsilon_{\alpha\beta\gamma}\nu^\alpha\wedge\nu^\beta\wedge\nu^\gamma
+\frac{1}{2}e^{-2g}h^{\alpha\delta}\epsilon_{\beta\gamma\delta}*\mathcal{D}A^I_\alpha\wedge\nu^\beta\wedge\nu^\gamma
\nonumber \\ & &
+\frac{1}{2}e^{-10g}\mathcal{F}^I_{\alpha\beta}h_{\gamma\delta}\epsilon_{\alpha\beta\delta}\omega_3\wedge\nu^\gamma,
\\
\hat{*}\hat{G}_3&=&-\frac{1}{3!}e^{3g}\tilde{h}\epsilon_{\alpha\beta\gamma}\nu^\alpha\wedge\nu^\beta\wedge\nu^\gamma
+\frac{1}{2}e^{4g}h^{\alpha\delta}\epsilon_{\beta\gamma\delta}*\bar{F}^\alpha\wedge\nu^\beta\wedge\nu^\gamma\nonumber
\\ & &
-\frac{1}{2}e^{-4g}\epsilon_{\alpha\beta\gamma}h_{\gamma\delta}*K_{\alpha\beta}\wedge\nu^\delta+\tilde{a}e^{-12g}\omega_3\,
\, .
\end{eqnarray}
The $\hat{*}$ and $*$ are Hodge dualities in six and three
dimensions, respectively. After using our ansatz in \eqref{G3eq},
\eqref{FIeq} and \eqref{THeq}, we find the following set of
equations
\begin{eqnarray}
& &\mathcal{D}(e^{2\theta+3g}\tilde{h})=0\, ,\label{eqDstarG13}\\
&
&\mathcal{D}(e^{\theta+6g}N^{\alpha\beta}*\bar{F}^\beta)+g_1c_1F^\alpha+\frac{1}{2}\epsilon_{\alpha\beta\gamma}N^{\alpha'\beta}N^{\beta'\gamma}
*K_{\alpha'\beta'}=0\, , \label{eqDstarG22}\\ & &
\mathcal{D}(N^{\alpha
\gamma}N^{\beta\delta}*K_{\alpha\beta})-g_1e^{\theta+6g}(N^{\alpha\gamma}*\bar{F}^\alpha\wedge
F^\delta-N^{\alpha\delta}*\bar{F}^\alpha\wedge F^\gamma)=0\,
,\label{eqDstarG31}\\ & &
\mathcal{D}(e^{\theta+6g}*\tilde{F}^I)+2c_1\tilde{F}^I-2e^{\theta+6g}N^{\alpha\alpha'}*\bar{F}^{\alpha'}\wedge\mathcal{D}A^I_\alpha  \nonumber\\
& &\qquad
+N^{\alpha\alpha'}N^{\beta\beta'}\mathcal{F}^I_{\alpha\beta}*K_{\alpha'\beta'}+g_2N^{\alpha\delta}f_{IJK}A^J_\delta*\mathcal{D}A^K_\alpha
=0 \, , \label{eqDstarF23}\\
&
&\mathcal{D}(N^{\alpha\beta}*\mathcal{D}A^I_\beta)+g_1e^{\theta+6g}*\tilde{F}^I\wedge
F^\alpha-2e^{\theta+6g}N^{\alpha\beta}*\bar{F}^\beta \wedge
\tilde{F}^I  \nonumber\\
&
&\qquad+2N^{\alpha\alpha'}N^{\beta\beta'}*K_{\alpha'\beta'}\wedge\mathcal{D}A^I_\beta+\frac{1}{2}e^{-\theta-6g}N^{\alpha'\beta}N^{\beta'\gamma}
\mathcal{F}^I_{\alpha'\beta'}\epsilon_{\alpha\beta\gamma}\omega_3
\nonumber\\ & &\qquad
-\tilde{a}e^{2\theta-12g}\epsilon_{\alpha\beta\gamma}\mathcal{F}^I_{\beta\gamma}\omega_3+g_2f_{IJK}e^{-\theta-6g}A^J_\beta\mathcal{F}^K_{\alpha'\beta'}
N^{\alpha'\beta}N^{\alpha\beta'}\omega_3 =0\, ,\label{eqDstarF32}
\\ & & 2d*dg-\frac{1}{3}d\ln
N+e^{\theta+6g}*\tilde{F}^I\wedge\tilde{F}^I+N^{\alpha\alpha'}*\mathcal{D}A^I_{\alpha'}\wedge\mathcal{D}A^I_\alpha
\nonumber \\
&
&\qquad+\frac{1}{2}e^{\theta+6g}N^{\alpha\alpha'}*\bar{F}^{\alpha'}\wedge\bar{F}^\alpha+\frac{1}{2}N^{\alpha\alpha'}N^{\beta\beta'}*K_{\alpha'\beta'}\wedge
K_{\alpha\beta}+ c_1^2e^{-2\theta-12g}\omega_3  \nonumber \\ & &
\qquad+\frac{1}{2}e^{-\theta-6g}N^{\alpha\alpha'}N^{\beta\beta'}\mathcal{F}^I_{\alpha\beta}\mathcal{F}^I_{\alpha'\beta'}\omega_3
+\tilde{a}^2e^{-12g}\omega_3=0\, , \label{thetaeqForm}
\end{eqnarray}
where we have used the summation convention on $\alpha, \beta,
\ldots$ regardless their upper or lower positions, and
$N^{\alpha\beta}\equiv(N^{-1})_{\alpha\beta}$. We have also used the
solution for equation \eqref{eqDstarG13} namely
\begin{equation}
\tilde{h}e^{2\theta+3g}=c_1
\end{equation}
with a constant $c_1$ in other equations. Equation
$\eqref{eqDstarG31}$ can be obtained by multiplying
\eqref{eqDstarF32} by $A^I_{\beta'}N^{\beta\beta'}$ and
antisymmetrizing in $\alpha$ and $\beta$. By using the explicit
forms of the Ricci tensors given in \cite{PopeSU2}, the scalar
equation \eqref{eqDstarF32}, after multiplied by
$e^{8g-\theta}N^{\beta\beta'}A^I_{\beta'}$ and symmetrized in
$\alpha$ and $\beta$, is the same as component $ij$ of the Einstein
equation with the trace part of Einstein equation taking care of
equation \eqref{thetaeqForm}. Component $ab$ the Einstein equation
will give three dimensional Einstein equation which we will not give
the explicit form here. Equation \eqref{eqDstarG22} gives Yang-Mills
equations for $A^\alpha$. The combination [\eqref{eqDstarF23}
$+2A^I_\alpha$\eqref{eqDstarG22}] gives Yang-Mills equations for
$A^I$
\begin{eqnarray}
& &\mathcal{D}[e^{\theta+6g}[(\delta_{IJ}+4A^I_\alpha A^J_\beta
N^{\alpha\beta})F^J-24g_1aA^I_\alpha
N^{\alpha\beta}F^\beta]]+2c_1F^I\nonumber \\ &
&+g_2f_{IJK}N^{\alpha\beta}A^J_\beta *\mathcal{D}A^K_\alpha
+g_2f_{IJK}N^{\alpha\alpha'}N^{\beta\beta'}A^J_\alpha A^K_\beta
*K_{\alpha'\beta'}=0\, .
\end{eqnarray}
We have checked that the equation for $F^\alpha$ is the same as
component $ai$ of the Einstein equation. So, there are two
Yang-Mills equations for $F^\alpha$ and $F^I$, one equation for $g$
and one equation for $A^I_\alpha$. All six dimensional field
equations are satisfied by our ansatz.
\subsection{Three dimensional gauged supergravity Lagrangian}
All three dimensional equations of motion obtained in the previous
subsection can be obtained from the following Lagrangian, with
$\hat{e}=ee^{3f+3g}=ee^{-6g}$,
\begin{eqnarray}
\mathcal{L}&=&\frac{1}{4}R*\mathbf{1}-\frac{1}{2}N^{-\frac{1}{3}}e^{8g}\big[(\delta_{IJ}+4A^I_\alpha
A^J_\beta N^{\alpha\beta})*F^I\wedge
F^J-48ag_1N^{\alpha\beta}A^I_\beta
*F^\alpha\wedge F^I\nonumber
\\ &
&+6ag_1^2(24aN^{\alpha\beta}-\delta_{\alpha\beta})*F^\alpha\wedge
F^\beta\big]-*d\big(2g-\frac{1}{12}\ln N\big)\wedge
d\big(2g-\frac{1}{12}\ln N\big) \nonumber
\\ & &
-\frac{1}{2}N^{\alpha\beta}*\mathcal{D}A^I_\alpha\wedge
\mathcal{D}A^I_\beta-N^{\alpha\alpha'}N^{\beta\beta'}A^I_\beta
A^J_{\beta'}*\mathcal{D}A^I_\alpha \wedge\mathcal{D}A^J_{\alpha'}
-V*\mathbf{1}+\mathcal{L}_{\textrm{CS}}\label{CSL}
\end{eqnarray}
which is the same as the dimensional reduction of the Lagrangian
\begin{equation}
\mathcal{L}_B=\frac{1}{4}\hat{R}\hat{*}\mathbf{1}-\frac{1}{4}\hat{*}\hat{d}\hat{\theta}\wedge\hat{d}\hat{\theta}-\frac{1}{2}e^{2\hat{\theta}}\hat{*}
\hat{G}_3\wedge\hat{G}_3-\frac{1}{2}e^{\hat{\theta}}\hat{*}\hat{F}^I\wedge\hat{F}^I
\end{equation}
together with the Chern-Simons terms. The scalar potential and the
Chern-Simons Lagrangian are given by
\begin{eqnarray}
V&=&\frac{1}{4}\big[N^{-\frac{2}{3}}(N_{\alpha\beta}N_{\alpha\beta}-\frac{1}{2}N_{\alpha\alpha}N_{\beta\beta})+2N^{-\frac{2}{3}}e^{-8g}\tilde{a}^2
\nonumber \\ & &
+N^{\frac{1}{3}}e^{-8g}N^{\alpha\alpha'}N^{\beta\beta'}\mathcal{F}^I_{\alpha\beta}\mathcal{F}^I_{\alpha'\beta'}-2c_1^2N^{\frac{2}{3}}e^{-16g}\big],
\\
\mathcal{L}_{\textrm{CS}}&=&2c_1\big(F^I\wedge
A^I-\frac{1}{6}g_2f_{IJK}A^I\wedge A^J\wedge A^K \big)\nonumber
\\ & &-12ag_1^2c_1\big(F^\alpha\wedge A^\alpha -\frac{1}{6}g_1\epsilon_{\alpha\beta\gamma}A^\alpha\wedge A^\beta\wedge
A^\gamma\big)\, .
\end{eqnarray}
In order to make formulae simpler and the symmetries of the scalar
manifold more transparent, we make the following rescalings. We
first restore the coupling $g_1$ in the appropriate places by
setting
\begin{equation}
a=\frac{\bar{a}}{g_1^2}.
\end{equation}
We can then remove the constant $a$ by setting
\begin{eqnarray}
c_1&=&\frac{\bar{c}_1}{6\bar{a}},\qquad
e^g=\frac{e^{\bar{g}}}{g_1^{\frac{1}{4}}}, \qquad
A^I_\alpha=\frac{\sqrt{6\bar{a}}}{g_1}\bar{A^I_\alpha},\nonumber \\
g_2&=& \frac{\bar{g}_2}{\sqrt{6\bar{a}}}, \qquad
N_{\alpha\beta}=\frac{6\bar{a}}{g_1^2}\bar{N}_{\alpha\beta}, \qquad
e^\theta=\frac{g_1^{\frac{3}{2}}}{6\bar{a}}e^{\bar{\theta}}\nonumber
\\ & & \textrm{and \qquad} A^I=\sqrt{6\bar{a}}\bar{A}^I\, .
\end{eqnarray}
After removing all the bars, we obtain the Lagrangian
\begin{eqnarray}
\mathcal{L}&=&\frac{1}{4}R*\mathbf{1}-\frac{1}{2}e^{2\sqrt{2}\Phi}[(\delta_{IJ}+4N^{\alpha\beta}A^I_\alpha
A^J_\beta)*F^I\wedge F^J-8N^{\alpha\beta}A^I_\beta *F^\alpha \wedge
F^I \nonumber \\ &
&+(4N^{\alpha\beta}-\delta_{\alpha\beta})*F^\alpha \wedge F^\beta]
-\frac{1}{2}*d\Phi \wedge d\Phi
-\frac{1}{2}N^{\alpha\beta}*\mathcal{D}A^I_\alpha \wedge
\mathcal{D}A^I_\beta \nonumber \\ & &
-N^{\alpha\alpha'}N^{\beta\beta'}A^I_\beta
A^J_{\beta'}*\mathcal{D}A^I_\alpha
\wedge\mathcal{D}A^J_{\alpha'}-V+\mathcal{L}_{\textrm{CS}}\label{YMLfinall}
\end{eqnarray}
where we have introduced the canonically normalized scalar for the
gauge singlet combination
\begin{equation}
\Phi=2\sqrt{2}g-\frac{\sqrt{2}}{12}\ln{N}\, .
\end{equation}
The scalar potential and Chern-Simons terms are now
\begin{eqnarray}
V&=&\frac{1}{4}\big[g_1^2N^{-1}e^{-2\sqrt{2}\Phi}(N_{\alpha\beta}N_{\alpha\beta}-\frac{1}{2}N_{\alpha\alpha}N_{\beta\beta})
+2N^{-1}e^{-2\sqrt{2}\Phi}\tilde{a}^2-2c_1^2e^{-4\sqrt{2}\Phi} \nonumber \\
& &
+e^{-2\sqrt{2}\Phi}N^{\alpha\alpha'}N^{\beta\beta'}\mathcal{F}^I_{\alpha\beta}\mathcal{F}^I_{\alpha'\beta'}\big]\label{potentialFinal},
\\
\mathcal{L}_{\textrm{CS}}&=&2c_1\bigg[F^I\wedge
A^I-\frac{1}{6}g_2f_{IJK}A^I\wedge A^J\wedge A^K \nonumber
\\ & &-\big(F^\alpha\wedge A^\alpha\wedge -\frac{1}{6}g_1\epsilon_{\alpha\beta\gamma}A^\alpha\wedge A^\beta\wedge
A^\gamma\big)\bigg] \label{KKPotential}
\end{eqnarray}
with
\begin{eqnarray}
\tilde{a}&=&g_1(1-A^I_\alpha A^I_\alpha)+\frac{1}{3}g_2A^3\, ,
\nonumber \\
\mathcal{F}^I_{\alpha\beta}&=&g_2A^JA^K
f_{IJK}-g_1\epsilon_{\alpha\beta\gamma}A^I_\gamma\,  .
\end{eqnarray}
We first look at the scalar matrix appearing in the gauge kinetic
terms
\begin{equation}
\mathbf{M}=\left(
              \begin{array}{cc}
                \mathcal{M}_{\alpha\beta} & \mathcal{M}_{\alpha J} \\
                \mathcal{M}_{I\beta} & \mathcal{M}_{IJ} \\
              \end{array}
            \right)
=e^{2\sqrt{2}\Phi}\left(
              \begin{array}{cc}
                4N^{\alpha\beta}-\delta_{\alpha\beta} & -4N^{\alpha\beta}A^J_\beta \\
                -4N^{\alpha\beta}A^I_\alpha & \delta_{IJ}+4N^{\alpha\beta}A^I_\alpha
A^J_\beta \\
              \end{array}
            \right).
\end{equation}
Introducing the matrix notation for
$A^I_{\phantom{a}\alpha}\equiv\mathbf{A}$ which is an $n\times 3$,
$n=\textrm{dim}G$, matrix and denoting
$\mathbb{I}_3=\mathbb{I}_{3\times 3}$ etc, we find
\begin{eqnarray}
N_{\alpha\beta}&\equiv &
\mathbf{N}=2\bigg(\mathbb{I}_3-\mathbf{A}^t\mathbf{A}\bigg),\qquad
N^{\alpha\beta}\equiv
\mathbf{N}^{-1}=\frac{1}{2\bigg(\mathbb{I}_3-\mathbf{A}^t\mathbf{A}\bigg)}, \\
\mathbf{M}&=& e^{2\sqrt{2}\Phi}\left(
                \begin{array}{cc}
                  \frac{\mathbb{I}_3+\mathbf{A}^t\mathbf{A}}{\mathbb{I}_3-\mathbf{A}^t\mathbf{A}}
                  & -2\frac{1}{\mathbb{I}_3-\mathbf{A}^t\mathbf{A}}\mathbf{A}^t \\
                  -2\frac{1}{\mathbb{I}_n-\mathbf{A}\mathbf{A}^t}\mathbf{A} &
                  \frac{\mathbb{I}_n
                  +\mathbf{A}^t\mathbf{A}}{\mathbb{I}_n-\mathbf{A}^t\mathbf{A}} \\
                \end{array}
              \right)\label{YMscalarmatrix}.
\end{eqnarray}
It follows that
\begin{equation}
\mathbf{M}^{-1}= e^{-2\sqrt{2}\Phi}\left(
                \begin{array}{cc}
                  \frac{\mathbb{I}_3+\mathbf{A}^t\mathbf{A}}{\mathbb{I}_3-\mathbf{A}^t\mathbf{A}}
                  & 2\frac{1}{\mathbb{I}_3-\mathbf{A}^t\mathbf{A}}\mathbf{A}^t \\
                  2\frac{1}{\mathbb{I}_n-\mathbf{A}\mathbf{A}^t}\mathbf{A} &
                  \frac{\mathbb{I}_n
                  +\mathbf{A}^t\mathbf{A}}{\mathbb{I}_n-\mathbf{A}^t\mathbf{A}} \\
                \end{array}
              \right).
\end{equation}
The scalars form the coset space $\mathbf{R}\times
\frac{SO(3,n)}{SO(3)\times SO(n)}$ with the factor $\mathbf{R}$
corresponding to $\Phi$. The scalar kinetic terms give rise to the
metric on $\mathbf{R}\times \frac{SO(3,n)}{SO(3)\times SO(n)}$
\begin{eqnarray}
& &-\frac{1}{2}*d\Phi \wedge d\Phi
-\frac{1}{2}N^{\alpha\beta}*\mathcal{D}A^I_\alpha \wedge
\mathcal{D}A^I_\beta -N^{\alpha\alpha'}N^{\beta\beta'}A^I_\beta
A^J_{\beta'}*\mathcal{D}A^I_\alpha \wedge\mathcal{D}A^J_{\alpha'}
\nonumber \\ & &=-\frac{1}{2}*d\Phi \wedge d\Phi
-\frac{1}{4}\textrm{Tr}\bigg(\frac{1}{\mathbb{I}_3-\mathbf{A}^t\mathbf{A}}*\mathcal{D}\mathbf{A}^t\wedge
\frac{1}{\mathbb{I}_n-\mathbf{A}\mathbf{A}^t}\mathcal{D}\mathbf{A}\bigg)\label{YMmetric}.
\end{eqnarray}
With all these results, the Lagrangian can be simply written as
\begin{eqnarray}
\mathcal{L}&=&\frac{1}{4}R*\mathbf{1}-\frac{1}{2}*d\Phi \wedge d\Phi
-\frac{1}{4}\textrm{Tr}\bigg(\frac{1}{\mathbb{I}_3-\mathbf{A}^t\mathbf{A}}*\mathcal{D}\mathbf{A}^t\wedge
\frac{1}{\mathbb{I}_n-\mathbf{A}\mathbf{A}^t}\mathcal{D}\mathbf{A}\bigg)\nonumber
\\ & &-\frac{1}{2}e^{2\sqrt{2}\Phi}\mathcal{M}_{\mathcal{A}\mathcal{B}}*F^{\mathcal{A}}\wedge
F^{\mathcal{B}}-V+\mathcal{L}_{\textrm{CS}}\label{YMLcomplete}
\end{eqnarray}
where $\mathcal{A}, \mathcal{B}=(\alpha, I)$.\\
 \indent We now come to supersymmetries of our truncated theory.
We will show that this truncation is indeed compatible with
supersymmetry namely supersymmetry transformations of various
components of $\hat{b}_{MN}$ must be consistent with our specific
choices of $C^\alpha=\frac{1}{2}A^I_\alpha A^I$. This ensures that
all the truncated fields will not be generated via supersymmetry.
For the filed $\hat{b}_{\mu\nu}$, we have eliminated it by using the
equation of motion for $\hat{G}_3$ in \eqref{eqDstarG13}. Because of
its non propagating nature, we do not need to worry about it. For
$\delta \hat{G}_{3\mu\alpha\beta}$ and $\delta
\hat{G}_{3\mu\nu\alpha}$, we have checked that they are consistent.
The detail of this check can be found in appendix \ref{detail}. We
have then verified that our truncated theory is a supersymmetric
theory. We will also give a confirmation to this claim in the next
section in which we will show that this theory is on-shell
equivalent to a manifestly supersymmetric $(SO(3)\ltimes
\mathbf{R}^3)\times (G\ltimes \mathbf{R}^n)$ Chern-Simons gauged
supergravity.
\\ \indent The final issue we should add here is the diagonalization
of the fermion kinetic terms. Applying the result of \cite{Pope3},
we find that our fermion kinetic Lagrangian can be written as
\begin{equation}
e^{-1}\mathcal{L}_{\textrm{Fkinetic}}=-\frac{1}{2}\bar{\psi}_\mu\Gamma^{\mu\nu\rho}D_\nu\psi_\rho-\frac{1}{2}\bar{\chi}\Gamma^\mu
D_\mu\chi -\frac{1}{2}(\delta^{IJ}+4N^{\alpha\beta}A^I_\alpha
A^I_\beta)\bar{\lambda}^I\Gamma^\mu \mathcal{D}_\mu \lambda^I
\end{equation}
where the three dimensional fields are given by
\begin{eqnarray}
\psi_a&=&e^{-\frac{3g}{2}}(\hat{\psi}_a+\Gamma_a
\Gamma^i\hat{\psi}_i)
\nonumber \\
\psi_i&=&e^{-\frac{3g}{2}}\big(\hat{\psi}_i-\frac{1}{2}\Gamma_i
\hat{\chi}\big)=2e^{-\frac{3g}{2}+\theta}A^I_i\hat{\lambda}^I,\,\, A^I_i=A^I_\alpha e^{-g}(L^{-1})^\alpha_i \nonumber \\
\chi &=&e^{-\frac{3g}{2}}\big(\Gamma^i
\hat{\psi}_i+\frac{1}{2}\hat{\chi}\big) \nonumber \\
\lambda^I&=&e^{\frac{\theta}{2}-\frac{3g}{2}}\hat{\lambda}^I\, .
\end{eqnarray}
\section{Chern-Simons and Yang-Mills gaugings in three
dimensions}\label{CSYM} In this section, we show the on-shell
equivalence between non-semisimple Chern-Simons and semisimple
Yang-Mills gaugings in three dimensions \cite{csym}. We will
construct Chern-Simons gauged supergravity with gauge groups
$(SO(3)\ltimes \mathbf{R}^3)\times (G\ltimes \mathbf{R}^n)$,
$n=\textrm{dim}G$, and show that the gauging is consistent according
to the criterion given in \cite{dewit}. We then show that this
theory is on-shell equivalent to the $SU(2)\times G$ gauged
supergravity obtained from $SU(2)$ reduction in the previous
section.
\\ \indent Before going to the discussion in details, we give
some necessary equations, we will use throughout this section, from
\cite{csym}. After integrate out the scalars and gauge fields
corresponding to the translations or shift symmetries, the
non-semisimple Chern-Simons gauged theory with scalar manifold $G/H$
becomes semisimple Yang-Mills gauged supergravity with smaller
scalar manifold $G'/H'$. The resulting Lagrangian is given by
\cite{csym}
\begin{eqnarray}
e^{-1}\tilde{\mathcal{L}}&=&\frac{1}{4}R+e^{-1}h_1\tilde{\mathcal{L}}_{\textrm{CS}}-\frac{1}{8}\mathbf{M}_{mn}F^{m\mu\nu}F^n_{\mu\nu}
-\frac{1}{4}G_{AB}\tilde{\mathcal{P}}^{A}_\mu\tilde{{\mathcal{P}}}^{B\mu}
\nonumber \\
& &
+\frac{1}{4}e^{-1}\epsilon^{\mu\nu\rho}\mathbf{M}_{mn}\tilde{\mathcal{V}}^{\underline{n}}_{\phantom{a}A}F^{m}_{\mu\nu}\tilde{\mathcal{P}}^{A}_\rho
-V \label{YML}
\end{eqnarray}
where all the notations are the same as that in \cite{csym} apart
from the metric signature, $(-++)$. We also use our notation for the
gauge couplings. We repeat here the quantities appearing in
\eqref{YML}
\begin{eqnarray}
G_{AB}&=&
\delta_{AB}-\tilde{\mathcal{V}}^{\underline{m}}_{\phantom{a}A}\mathbf{M}_{mn}\tilde{\mathcal{V}}^{\underline{n}}_{\phantom{a}B},\qquad
\mathbf{M}_{mn}=(\tilde{\mathcal{V}}^{\underline{m}}_{\phantom{a}A}\tilde{\mathcal{V}}^{\underline{n}}_{\phantom{a}A})^{-1},
\nonumber \\
 \tilde{\mathcal{L}}_{\textrm{CS}}&=&\frac{1}{4}\epsilon^{\mu\nu\rho}A_\mu
 ^{m}\eta_{mn}\big(\pd_\nu A^n_\rho+\frac{1}{3}g_1 f^n_{\phantom{s}kl}A^k_\nu
 A^l_\rho\big) \nonumber \\ & &
+\frac{1}{4}\epsilon^{\mu\nu\rho}A_\mu
 ^m\eta_{mn}\big(\pd_\nu A^n_\rho+\frac{1}{3}g_2 f^n_{\phantom{a}kl}A^k_\nu
 A^l_\rho\big),
 \nonumber \\
 \tilde{\mathcal{V}}^{\mathcal{M}}_{\phantom{a}\mathcal{A}}t^{\mathcal{A}}&=&\tilde{L}^{-1}t^{\mathcal{M}}\tilde{L},
 \nonumber \\
 \tilde{\mathcal{Q}}_\mu+\tilde{\mathcal{P}}_\mu &=&
 \tilde{L}^{-1}(\pd_\mu +g_1\eta_{1mn}A^m_{1\mu}
 t_1^n+g_2\eta_{2mn}A^m_{2\mu}
 t_2^n)\tilde{L}\, .\label{YMLQ}
\end{eqnarray}
\indent We are now in a position to construct a consistent
Chern-Simons gauged supergravity with gauge groups $(SO(3)\ltimes
\mathbf{R}^3)\times (G\ltimes \mathbf{R}^n)$. We proceed as in
\cite{gkn} using the formulation of \cite{dewit}.
\\ \indent The $4(1+n)$ scalar fields are described by a coset space
$\frac{SO(4,1+n)}{SO(4)\times SO(n+1)}$. We parametrize the coset by
\begin{equation}
L=\left(
    \begin{array}{cc}
      A & B \\
      B^t & C \\
    \end{array}
  \right)
\end{equation}
where $A$ is a symmetric $4\times 4$ matrix, $B$ is a $4\times
(n+1)$ matrix, and $C$ is a symmetric $(n+1)\times (n+1)$ matrix.
These matrices satisfy the relations
\begin{eqnarray}
A^2-BB^t&=&\mathbb{I}_4, \nonumber \\
AB-BC&=&0,\nonumber \\
C^2-B^tB&=&\mathbb{I}_{n+1}\, .
\end{eqnarray}
The gauging is characterized by the embedding tensor
\begin{equation}
\Theta_{\mathcal{M}\mathcal{N}}
=g_1\delta_{a_1b_1}+g_2\delta_{a_2b_2}+h_1\delta_{b_1b_1}+h_2\delta_{b_2b_2}.
\end{equation}
The ranges of the indices are $a_1, b_1=1,2,3$ and $a_2,
b_2=1,\dots, n$. We denote the $(5+n)\times (5+n)$ matrix in the
block form
\begin{equation}
\left(
   \begin{array}{c|c}
     4\times 4 & 4\times (n+1)   \\ \hline
     (n+1)\times 4 & (n+1)\times (n+1)
   \end{array}
 \right).
\end{equation}
With this form, the generators of $SO(4,1+n)$ can be shown as
\begin{equation}
\left(
   \begin{array}{c|c}
     J_{SO(4)} & Y   \\ \hline
     Y^t & J_{SO(n+1)}
   \end{array}
 \right)
\end{equation}
with $Y$ being non-compact and given by $e_{a\hat{I}}+e_{\hat{I}a}$.
We further divide each block by separating its last row and last
column from the rest and use the following ranges of indices:
\begin{displaymath}
\alpha, \, \beta=1, 2, 3, \qquad I, \, J=1,\ldots, n , \qquad
\hat{I}=5,\ldots, n+5,\,\, \textrm{and} \,\, a,\, b=1,\ldots , 4.
\end{displaymath}
Various gauge groups are described by the following generators:
\begin{eqnarray}
  SO(3)&: & J^\alpha_{a_1}= \epsilon_{\alpha\beta\gamma}
  e_{\beta\gamma}, \nonumber \\
  G &:& J^I_{a_2}=f^I_{\phantom{a}JK}e_{JK}, \nonumber \\
  \mathbf{R}^3
  &:&J^\alpha_{b_1}=e_{\alpha,n+5}+e_{n+5,\alpha}+e_{4\alpha}-e_{\alpha
  4},\nonumber \\
  \mathbf{R}^n
  &:&J^I_{b_2}=e_{4,I+4}+e_{I+4,4}+e_{n+5,I+4}-e_{I+4,n+5}\nonumber \\
  & & \textrm{with}\,\,\, (e_{ab})_{cd}=\delta_{ac}\delta_{bd},\, \textrm{etc}\,. \label{gauge generator}
\end{eqnarray}
Schematically, these gauge generators are embedded in the
$(5+n)\times (5+n)$ matrix as
\begin{equation}
\left(
   \begin{array}{cc|c|cccc|c}
      & J_{a_1}(3\times 3) & -b_1 &  &  &  &  & b_1 \\
      &  & (3\times 1) &  &  &  &  & (3\times 1) \\ \hline
      &  b_1^t (1\times 3) & &  & b_2^t (1\times n) &  &  \\ \hline
      &  &  &  &  &  &  &  \\
      &  &   b_2&  &   J_{a_2} & &  & -b_2 \\
      &  & (n\times 1) &  & (n\times n) & &  & (n\times 1) \\
      &  &  &  &  &  &  &  \\ \hline
      &  b_1^t(1\times 3)&  &  &  b_2^t (1\times n) & &  &  \\
   \end{array}
 \right)
\end{equation}
where each $b_1$ and $b_2$ correspond to various $e$'s factors in
$J_{b_1}$ and $J_{b_2}$ in \eqref{gauge generator}. Notice that the
shift generators have components in both $SO(4)\times SO(n+1)$ and
$Y$ parts. Furthermore, $J_{b_1}$ and $J_{b_2}$ transform as adjoint
representations of the gauge groups $SO(3)$ and $G$, respectively.
\\ \indent From this information, we can construct T-tensors and
check the consistency of the gauging according to the criterion
given in \cite{dewit}, $\mathbb{P}_\boxplus T^{IJ,KL}=0$. The
consistency requires that
\begin{equation}
h_2=-h_1\, .
\end{equation}
The  $4(1+n)$ scalars correspond to the non-compact generators $Y$.
After using the shift symmetries to remove some of the shifted
scalars and gauge fields, we are left with $1+3n$ scalars embedded
in $(5+n)\times (5+n)$ matrix as
\begin{equation}
\tilde{L}=\left(
   \begin{array}{cc|c|cccc|c}
      &\frac{1}{\sqrt{\mathbb{I}_{3}-\mathbf{A}^t\mathbf{A}}}  &  &  & &\frac{1}{\sqrt{\mathbb{I}_{3}-\mathbf{A}^t\mathbf{A}}}\mathbf{A}^t &    &  \\
      &  &  &  &  &  &  &  \\ \hline
      &  & \cosh{\sqrt{2}\Phi} &  &  &  & & \sinh{\sqrt{2}\Phi} \\ \hline
      &  &  &  &  &  &  &  \\
      & \frac{1}{\sqrt{\mathbb{I}_{n}-\mathbf{A}\mathbf{A}^t}}\mathbf{A} &   &  &  &  \frac{1}{\sqrt{\mathbb{I}_{n}-\mathbf{A}\mathbf{A}^t}} & & \\
      &  &  &  &  & &  &  \\
      &  &  &  &  &  &  &  \\ \hline
      &  & \sinh{\sqrt{2}\Phi} &  &   & &  & \cosh{\sqrt{2}\Phi} \\
   \end{array}
 \right)\label{coset1}.
\end{equation}
Note that in \eqref{coset1}, we have chosen a specific form of $A$,
$B$ and $C$. $\mathbf{A}$ is an $n\times 3$ matrix to be identified
with $A^I_\alpha$ in the previous section. The resulting coset space
is readily recognized as $\mathbf{R}\times
\frac{SO(3,n)}{SO(3)\times SO(n)}$ in which $\Phi$ corresponds to
the $\mathbf{R}\sim SO(1,1)$ part.
\\
\indent In this $(SO(3)\ltimes \mathbf{R}^3)\times(G\ltimes
\mathbf{R}^n)$ gauged theory, we find that
\begin{equation}
\tilde{\mathcal{V}}^{\underline{n}}_{\phantom{a}A}\tilde{\mathcal{P}}^A_\mu=0\,
\end{equation}
by using $\mathcal{V}$'s given in appendix \ref{Appendix B} and
computing $\tilde{\mathcal{P}}^A$ from \eqref{YMLQ}. So, there is no
coupling term between scalars and gauge field strength in
\eqref{YML}. Another consequence of this is that the scalar metric
$G_{AB}$ in \eqref{YML} is effectively $\delta_{AB}$.
\\ \indent From \eqref{coset},
we can compute the scalar manifold metric which is given by the
general expression
\begin{equation}
ds^2=\frac{1}{8}\textrm{Tr}(\tilde{L}^{-1}d\tilde{L}|_Y\tilde{L}^{-1}d\tilde{L}|_Y)
\end{equation}
where $|_Y$ means that we take the coset component of the
corresponding one-form. Using the relation
$\mathbf{A}^t\frac{1}{\sqrt{\mathbb{I}_{n}-\mathbf{A}\mathbf{A}^t}}=\frac{1}{\sqrt{\mathbb{I}_{3}-\mathbf{A}^t\mathbf{A}}}\mathbf{A}^t$,
we find, after a straightforward calculation,
\begin{equation}
\tilde{L}^{-1}d\tilde{L}|_Y=\left(
   \begin{array}{cc|c|cccc|c}
      &0  &  &  & &\frac{1}{\sqrt{\mathbb{I}_{3}-\mathbf{A}^t\mathbf{A}}}d\mathbf{A}^t\frac{1}{\sqrt{\mathbb{I}_{n}-\mathbf{A}\mathbf{A}^t}} &    &  \\
      &  &  &  &  &  &  &  \\ \hline
      &  & 0 &  &  &  & & \sqrt{2}d\Phi \\ \hline
      &  &  &  &  &  &  &  \\
      & \frac{1}{\sqrt{\mathbb{I}_{n}-\mathbf{A}\mathbf{A}^t}}d\mathbf{A}\frac{1}{\sqrt{\mathbb{I}_{3}-\mathbf{A}^t\mathbf{A}}} &   &  &  &  0 & & \\
      &  &  &  &  & &  &  \\
      &  &  &  &  &  &  &  \\ \hline
      &  & \sqrt{2}d\Phi &  &   & &  & 0 \\
   \end{array}
 \right)
\end{equation}
where we have given only the coset components to simplify the
equation. The scalar metric is then given by
\begin{equation}
ds^2=\frac{1}{2}d\Phi d\Phi
+\frac{1}{4}\textrm{Tr}\bigg(\frac{1}{\mathbb{I}_3-\mathbf{A}^t\mathbf{A}}d\mathbf{A}^t
\frac{1}{\mathbb{I}_n-\mathbf{A}\mathbf{A}^t}d\mathbf{A}\bigg).
\end{equation}
This is exactly the same scalar metric appearing in the scalar
kinetic terms in \eqref{YMmetric}. The scalar matrix appearing in
the gauge field kinetic terms can be computed as follows. From
\eqref{YML} and \eqref{YMLQ}, we can write
\begin{equation}
\mathbf{M}_{mn}=\bar{\mathbf{M}}^{-1}_{mn}\qquad \textrm{where}
\qquad
\bar{\mathbf{M}}_{mn}=\tilde{\mathcal{V}}^{\underline{m}}_{\phantom{a}A}\tilde{\mathcal{V}}^{\underline{n}}_{\phantom{a}A}\,
.
\end{equation}
In our case, the indices $\underline{m}, \underline{n}=b_1, b_2$,
and $m,n=a_1,a_2$. With properly normalized coset generators $Y^A$,
we find that
\begin{eqnarray}
\bar{\mathbf{M}}&=&\left(
          \begin{array}{cc}
            \bar{\mathbf{M}}_{a_1a_1} & \bar{\mathbf{M}}_{a_1a_2} \\
            \bar{\mathbf{M}}_{a_2a_1} & \bar{\mathbf{M}}_{a_2a_2} \\
          \end{array}
        \right),\qquad
        \bar{\mathbf{M}}_{a_{\hat{i}}a_{\hat{j}}}=\tilde{\mathcal{V}}^{b_{\hat{i}}}_{\phantom{a}A}\tilde{\mathcal{V}}^{b_{\hat{j}}}_{\phantom{a}A}
        \nonumber \\
        \textrm{where}\,\, \hat{i}, \, \hat{j}&=&1,2\,\, \textrm{and}\,\,
        \tilde{\mathcal{V}}^{b_{\hat{i}}}_{\phantom{a}A}=\textrm{Tr}(\tilde{L}^{-1}J_{b_{\hat{i}}}\tilde{L}Y^A).
\end{eqnarray}
After some algebra, we find that the matrix
$\mathbf{M}_{a_{\hat{i}}a_{\hat{j}}}$ is the same as
$\mathcal{M}_{\mathcal{A}\mathcal{B}}$ in \eqref{YMLcomplete}. So,
the reduced scalar coset from the Chern-Simons gauged theory is the
same as that in the Yang-Mills gauged theory obtained form the
$SU(2)$ reduction.
\\ \indent
Finally, we have to check the scalar potential. From the embedding
tensor, we can compute the potential by using \cite{dewit}
\begin{equation}
V=A_1^{\bar{I}\bar{J}}A_1^{\bar{I}\bar{J}}-2g^{ij}A_{2i}^{\bar{I}\bar{J}}A_{2j}^{\bar{I}\bar{J}}.\label{CSV}
\end{equation}
We find that potential obtained here is exactly the same as in
\eqref{potentialFinal}. We give the details of this calculation in
appendix \ref{Appendix B}. We have now completely shown that the
Chern-Simons gauged theory constructed in this section is the same
as the Yang-Mills gauged theory obtained from the $SU(2)$ reduction
in the previous section.
\section{Conclusions}\label{conclusions}
We have obtained Yang-Mills $SU(2)\times G$ gauged supergravity in
three dimensions from $SU(2)$ group manifold reduction of six
dimensional (1,0) supergravity coupled to an anti-symmetric tensor
and $G$ Yang-Mills multiplets. We have also given consistent
truncations in both bosonic and fermionic fields from which the
resulting consistent reduction ansatz followed. The truncation,
which removes three dimensional massive vector fields, results in an
$N=4$ supergravity theory describing $4(1+\textrm{dim}\, G)$ bosonic
propagating degrees of freedom, $1+3\textrm{dim}\, G$ scalars and
$3+\textrm{dim}\, G$ gauge fields, together with $4(1+\textrm{dim}\,
G)$ fermions. The scalar fields are coordinates in the coset space
$\mathbf{R}\times \frac{SO(3,\, \textrm{dim}\, G)}{SO(3)\times
SO(\textrm{dim}\, G)}$.
 \\
\indent Furthermore, we have explicitly constructed the $N=4$
Chern-Simons $(SO(3)\ltimes \mathbf{R}^3)\times (G\ltimes
\mathbf{R}^{\textrm{dim}\, G})$ gauged supergravity in three
dimensions, following the general procedure detailed in
\cite{dewit}. The scalar manifold $\frac{SO(4,\, 1+\textrm{dim}\,
G)}{SO(4)\times SO(1+\textrm{dim}\, G)}$ becomes
$\mathbf{R}\times\frac{SO(3,\, \textrm{dim}\, G)}{SO(3)\times
SO(\textrm{dim}\, G)}$ after removing the scalars corresponding to
the translations or shift symmetries. We have shown the agreement
between the resulting Lagrangian and the Lagrangian obtained from
dimensional reduction i.e. the gauge field kinetic terms, the scalar
manifold metrics and scalar potentials.
\\ \indent
We have not given the supersymmetry transformations of the three
dimensional fields here. These can, in principle, be obtained by
direct computations or using the results in \cite{Pope3} with our
truncations. Although supersymmetry transformations of fermions are
essential, for example  for finding BPS solutions, it is more
convenient to work with the equivalent Chern-Simons gauged theory as
the latter turns out to be simpler than the equivalent Yang-Mills
theory, see \cite{csym} for a discussion. In particular, the
consistency of the Chern-Simons gauging is encoded in a single
algebraic condition on the embedding tensor \cite{nicolai1,nicolai2,
nicolai3, dewit}.
\\ \indent
In the case where $G=SU(2)$, the $SU(2)\times SU(2)$ Yang-Mills
gauged theory is the same as $(SU(2)\ltimes \mathbf{R}^3)^2$
Chern-Simons gauged theory with scalar manifold $\frac{SO(4,\,
4)}{SO(4)\times SO(4)}$. Such quaternionic space has been considered
in \cite{gkn}, however the $(SU(2)\ltimes \mathbf{R}^3)^2$ gauging
appearing there is different from the one in this paper. The two
gauged $SU(2)$'s in \cite{gkn} are the diagonal subgroups of the two
$SU(2)_L$ and the two $SU(2)_R$ respectively of the $SO(4)\times
SO(4)$. The latter can be constructed using the parametrization of
the target space in terms of $e$ and $B$ matrices as in \cite{gkn}.
The action of shift symmetry generators is to shift $B$. We can
simply set $B=0$ in this parametrization to obtain the Yang-Mills
coset. Although the identification of $(A^I_\alpha,\Phi)$ and $e$ is
complicated, with the help of \textsl{Mathematica},
it can be shown that the two theories are indeed equivalent. \\
\indent It is interesting to study RG flow solutions in both
Chern-Simons and Yang-Mills gauged theories, which, by the present
results, can be lifted up to six dimensions. We will report the
results of this analysis in a forthcoming paper \cite{new}. A
natural open problem is how to obtain  3D $N=4$ gauged supergravity
with two quaternionic scalar manifolds, for example to recover the
theory studied in \cite{gkn}. Presumably we would need to add
hypermultiplets to the six dimensional theory, whose scalars
themselves live on a quaternionic manifold, or perhaps, we may even
need to start with extended supersymmetry in six dimensions.

\acknowledgments This work has been supported in part by the EU
grant UNILHC-Grant Agreement PITN-GA-2009-237920.

\appendix
\section{Details of the calculations}\label{detail}
In this appendix, we present some details of the calculations
mentioned in section \ref{reduction}.
\subsection{Supersymmetry of the fermionic truncation}
In order to check the supersymmetry transformation of
\eqref{fermiontruncation}, we start by putting our ansatz to the
$\delta\hat{\psi}_i$, $\delta \hat{\chi}$ and
$\delta\hat{\lambda}^I$ given in \eqref{6Dsusyvar}. The result is
\begin{eqnarray}
\delta \hat{\psi}_i &=&\frac{1}{8}g_1e^{g-2f}{F\Slash}^i (1\otimes
1\otimes 1)\epsilon
-\frac{i}{2}e^{-f}({P\Slash}_{ij}-{\pd\Slash}\,g\delta_{ij})(1\otimes
\gamma^j\otimes
\sigma_3)\epsilon\nonumber \\
&
&+\frac{i}{2}e^{-g}\big(T_{ij}-\frac{1}{2}T\delta_{ij}\big)(1\otimes
\gamma^j\otimes 1)\epsilon
+e^\theta\big[\frac{1}{8}e^{-2f-g}(L^{-1})^\alpha_j{\bar{F}\Slash}^\alpha(1\otimes
\gamma^j\otimes \sigma_2)
\nonumber \\
& &+\frac{1}{4}\tilde{h}(1\otimes 1\otimes \sigma_1)
+\frac{i}{4}e^{-f-2g}(L^{-1})^\beta_l(L^{-1})^\gamma_j\epsilon_{ljk}A^I_\gamma{\mathcal{D}\Slash}A^I_\beta(1\otimes
\gamma^k\otimes \sigma_1)\nonumber \\
& &+\frac{i}{4}\tilde{a}e^{-3g}(1\otimes 1\otimes \sigma_2)
\big](1\otimes \gamma^i\otimes
\sigma_2)\epsilon\label{deltapsiI}\\
\delta\hat{\chi}&=&\frac{1}{2}{\pd\Slash}\,\theta
\epsilon-e^\theta\big[\frac{1}{2}\tilde{h}(1\otimes 1\otimes
\sigma_1)+\frac{1}{4}e^{-2f-g}(L^{-1})^\alpha_i{\bar{F}\Slash}^\alpha(1\otimes
\gamma^i\otimes \sigma_2)\nonumber \\
& &
+\frac{i}{2}e^{-f-2g}(L^{-1})^\beta_i(L^{-1})^\gamma_j\epsilon_{ijk}A^I_\gamma{\mathcal{D}\Slash}A^I_\beta(1\otimes
\gamma^k\otimes \sigma_1)\nonumber \\
& &+\frac{i}{2}\tilde{a}e^{-3g}(1\otimes
1\otimes \sigma_2) ]\epsilon\label{deltachi}\\
\delta\hat{\lambda}^I&=&\frac{1}{4}[e^{-2f}{\tilde{F}\Slash}^I(1\otimes
1\otimes
1)+2ie^{-f-g}(L^{-1})^\alpha_i{\mathcal{D}\Slash}A^I_\alpha(1\otimes
\gamma^i\otimes \sigma_3)\nonumber \\
& &
+ie^{-2g}(L^{-1})^\alpha_i(L^{-1})^\beta_j\epsilon_{ijk}\mathcal{F}^I_{\alpha\beta}(1\otimes
\gamma^k\otimes 1) ]\epsilon\label{deltalambda}
\end{eqnarray}
We have used the notations
${\hat{F}\Slash}^I=\hat{F}^I_{MN}\Gamma^{MN}$ etc. From these
equations and
$\mathbb{I}_2\otimes\mathbb{I}_2\otimes\sigma_3\epsilon^A=\epsilon^A$,
we find, up to leading order in fermions, that
\begin{equation}
\delta\hat{\psi}_i-\frac{1}{2}\Gamma_i\delta\hat{\chi}-2e^{\theta-g}A^I_\alpha
(L^{-1})^\alpha_i\delta\hat{\lambda}^I=0
\end{equation}
provided that
\begin{equation}
h_{\alpha\beta}=e^{\theta-2g}(12a\delta_{\alpha \beta}-2A^I_\alpha
A^I_\beta)\equiv e^{\theta-2g}N_{\alpha\beta}.
\end{equation}
In proving this result, the following relations are useful
\begin{eqnarray}
L^i_\alpha L^j _\beta
T_{ij}&=&e^{2\theta-4g}N_{\alpha\gamma}N_{\beta\gamma} \nonumber \\
T&=&T_{ii}=e^{\theta-2g}N_{\alpha\alpha} \nonumber \\
L^i_\alpha L^j_\beta
P_{aij}&=&\frac{1}{2}D_a(e^{\theta-2g}N_{\alpha\beta}).
\end{eqnarray}
\subsection{Supersymmetry of the bosonic truncation}
To check \eqref{deltah}, we start by noting that
\begin{equation}
\hat{g}_{\alpha\beta}=e^{2g}h_{\alpha\beta}=e^\theta(12a\delta_{\alpha\beta}-2A^I_\alpha
A^I_\beta).
\end{equation}
It follows that, with
$\hat{g}_{\alpha\beta}=\frac{1}{2}(\Gamma_\alpha\Gamma_\beta+\Gamma_\beta\Gamma_\alpha)$,
\begin{eqnarray}
\delta \hat{g}_{\alpha\beta}&=&\delta\theta
\hat{g}_{\alpha\beta}-2e^\theta(A^I_\alpha \delta A^I_\beta+\delta
A^I_\alpha
 A^I_\beta),\, \, \textrm{or}\nonumber \\
\bar{\epsilon}(\Gamma_\alpha \psi
_\beta+\Gamma_\beta\psi_\alpha)&=&\delta\theta
\hat{g}_{\alpha\beta}-2e^\theta (A^I_\alpha \delta
A^I_\beta+A^I_\beta
\delta A^I_\alpha)\nonumber \\
&=&\bar{\epsilon}\frac{1}{2}(\Gamma_\alpha\Gamma_\beta+\Gamma_\beta
\Gamma_\alpha)\chi +2e^\theta
\bar{\epsilon}(\Gamma_\beta\lambda^IA^I_\alpha+\Gamma_\alpha
\lambda^IA^I_\beta),\nonumber \\
\textrm{or} \,\,\, & &
\bar{\epsilon}\Gamma_\alpha\bigg(\psi_\beta-\frac{1}{2}\Gamma_\beta\chi-2e^\theta
A^I_\beta\lambda^I\bigg)+(\alpha\leftrightarrow \beta)=0
\end{eqnarray}
where we have temporarily dropped the hats on the fermions in order
to simplify the equations. \\ \indent We then move to the
supersymmetry transformations of $\hat{b}_{MN}$. It is more
convenient to work with the transformation of the filed strength
$\hat{G}_3$. With equation \eqref{fermiontruncation}, the component
$\delta\hat{b}_{\alpha\beta}$ vanishes identically. The
$\delta\hat{G}_{3\mu \alpha\beta}$ gives the condition
\begin{equation}
\delta\hat{b}_{\mu\alpha}=\delta (A^I_\alpha
A^I_\mu-6ag_1A^\alpha_\mu)\, .
\end{equation}
Using $A^I_\mu=\hat{A}^I_\mu+g_1A^I_\alpha A^\alpha_\mu$ and $\delta
\hat{b}_{\mu\alpha}$ from \eqref{6Dsusyvar}, we find that
\begin{eqnarray}
\delta\hat{b}_{\mu\alpha}-\delta (A^I_\alpha
A^I_\mu-6ag_1A^\alpha_\mu)&=&
 2A^I_\alpha
 \bar{\epsilon}\Gamma_\mu\hat{\lambda}^I-e^{-\theta}\bar{\epsilon}\Gamma_\mu\hat{\psi}_\alpha+\frac{1}{2}e^{-\theta}\bar{\epsilon}(\Gamma_{\mu\alpha}
 +\hat{g}_{\mu\alpha})\hat{\chi} \nonumber \\ &=&
 -e^{-\theta}\bar{\epsilon}\Gamma_\mu\big(\hat{\psi}_\alpha-\frac{1}{2}\Gamma_\alpha
 \hat{\chi}-2e^{\theta}A^I_\alpha\hat{\lambda}^I\big)=0
\end{eqnarray}
where we have used
\begin{equation}
\hat{g}_{\mu\alpha}=\hat{e}^i_\mu
\hat{e}^i_\alpha=-g_1h_{\alpha\beta}e^{2g}A^\beta_\mu=-g_1e^{\theta}N_{\alpha\beta}A^\beta_\mu\,
.
\end{equation}
Note that
\begin{equation}
\hat{\psi}_i-\frac{1}{2}\Gamma_i\hat{\chi}-2e^{\theta-g}A^I_\alpha
(L^{-1})^\alpha_i\hat{\lambda}^I=e^{-g}(L^{-1})^\alpha_i
\big(\hat{\psi}_\alpha-\frac{1}{2}\Gamma_\alpha\hat{\chi}-2e^\theta
A^I_\alpha\hat{\lambda}^I\big).
\end{equation}
$\delta \hat{G}_{3\mu\nu\alpha}$ is simply the derivative of the
previous result namely
\begin{equation}
\delta \hat{G}_{3\mu\nu\alpha}=2\pd_{[\mu}\delta
\hat{b}_{\nu]\alpha}=2\pd_{[\mu}\delta(A^I_{\nu]}A^I_\alpha-6ag_1A^\alpha_{\nu]}).
\end{equation}
\section{Essential formulae for $N=4$, $(SO(3)\ltimes \mathbf{R}^3)\times (G\ltimes \mathbf{R}^n)$ gauged
supergravity}\label{Appendix B} In this appendix, we give some
details of the calculation and needed quantities in the $N=4$,
$(SO(3)\ltimes \mathbf{R}^3)\times (G\ltimes \mathbf{R}^n)$ gauged
supergravity constructed in section \ref{CSYM}. We recall here the
useful expressions for coset space
\begin{eqnarray}
L^{-1} D_\mu L&=& \frac{1}{2}Q^{\bar{I}\bar{J}}_\mu
X^{\bar{I}\bar{J}}+Q^\alpha_\mu
X^{\alpha}+e^A_\mu Y^A\, , \nonumber \\
L^{-1}t^\mathcal{M}L&=&\frac{1}{2}\mathcal{V}^{\mathcal{M}\bar{I}\bar{J}}X^{\bar{I}\bar{J}}+\mathcal{V}^\mathcal{M}_{\phantom{a}\alpha}X^\alpha+
\mathcal{V}^\mathcal{M}_{\phantom{a}A}Y^A.\label{coset}
\end{eqnarray}
The tensors $A_1$ and $A_2$ can be obtained from the T-tensors by
using, with $N=4$,
\begin{eqnarray}
A_1^{\bar{I}\bar{J}}&=&-\frac{4}{N-2}T^{\bar{I}\bar{M},\bar{J}\bar{M}}+\frac{2}{(N-1)(N-2)}\delta^{\bar{I}\bar{J}}T^{\bar{M}\bar{N},\bar{M}
\bar{N}},\nonumber\\
A_{2j}^{\bar{I}\bar{J}}&=&\frac{4}{N(N-2)}f^{\bar{M}(\bar{I}
m}_{\phantom{as}j}T^{\bar{J})\bar{M}}_{\phantom{as}m}+\frac{2}{N(N-1)(N-2)}\delta^{\bar{I}\bar{J}}f^{\bar{K}\bar{L}\phantom{a}m}_{\phantom{as}j}T^{
\bar{K}\bar{L}}_{\phantom{as}m}
+\frac{2}{N}T^{\bar{I}\bar{J}}_{\phantom{as}j}
\end{eqnarray}
where $\bar{I}, \bar{J}, \ldots=1,\ldots 4$ label the R-symmetry
indices. The coordinate index on the target space $i$ will be
denoted by a pair of indices specifying the entries of the $L$. In
order to simplify the equations, we introduce a symbolic notation
$R$ for the R-symmetry generators including their indices. We start
by giving all the
$\mathcal{V}^{\mathcal{M}}_{\phantom{a}\bar{I}\bar{J}}$'s.
\begin{eqnarray}
\mathcal{V}^R_{a_\alpha}&=&\frac{1}{4}\epsilon_{\alpha\beta\gamma}(ARA)_{\gamma\beta},
\qquad a_\alpha=\epsilon_{\alpha\beta\gamma}e_{\beta\gamma},
\nonumber \\
\mathcal{V}^R_{a_I}&=&-\frac{1}{4}f^I_{\phantom{a}JK}(B^tRB)_{JK},\qquad
\mathcal{V}^R_{b_\alpha}=\frac{1}{2}\mathcal{H}(AR)_{\alpha 4},
\,\,\,
\mathcal{H}=A_{44}-B_{4,n+1}, \nonumber \\
\mathcal{V}^R_{b_I}&=&\frac{1}{2}\mathcal{H}(B^tR)_{I4}\, .
\end{eqnarray}
The $\mathcal{V}^{\mathcal{M}}_{\phantom{a}i}$'s are given by
\begin{eqnarray}
\mathcal{V}^{a_\alpha}_{\delta
L}&=&\epsilon_{\alpha\beta\gamma}B_{\gamma L}, \qquad
\mathcal{V}^{a_I}_{\delta M} =-f_{IJK}B_{\delta J}C_{KM}, \qquad
\mathcal{V}^{b_\alpha}_{\delta,n+1}=\mathcal{H}A_{\delta \alpha}, \nonumber \\
\mathcal{V}^{b_\alpha}_{4L}&=&\mathcal{H}B_{\alpha L}, \qquad
\mathcal{V}^{b_I}_{4L}= \mathcal{H}C_{IL}, \qquad
\mathcal{V}^{b_I}_{\delta ,n+1}=\mathcal{H}B_{\delta I}.
\end{eqnarray}
The T-tensors are defined by
\begin{equation}
T_{\mathcal{A}\mathcal{B}}=\Theta_{\mathcal{M}\mathcal{N}}\mathcal{V}^{\mathcal{M}}_{\phantom{a}\mathcal{A}}
\mathcal{V}^{\mathcal{N}}_{\phantom{a}\mathcal{B}}
\end{equation}
which gives
\begin{eqnarray}
T^{RR'}&=&\frac{1}{16}\big[-8g_1R_{\alpha 4}R'_{\alpha
4}\textrm{det}A+8g_2\mathcal{H}\frac{B^3}{6}+4h_1\mathcal{H}^2R_{\gamma
4}R'_{\gamma 4}\big],
\nonumber \\
T^R_{\delta,
n+1}&=&\frac{1}{4}\big[-2g_1\mathcal{H}\textrm{det}AR_{\delta
4}+2g_2\mathcal{H}\frac{B^3}{6}R_{\delta 4}+2h_1\mathcal{H}^2R_{\delta 4}\big],\nonumber \\
T^R_{4
L}&=&\frac{1}{4}\big[-g_1\mathcal{H}\epsilon_{\alpha\beta\gamma}B_{\alpha
L}(ARA)_{\beta\gamma}+g_2\mathcal{H}f^I_{\phantom{a}JK}C_{IL}(B^tRB)_{JK}\big] ,\nonumber \\
T^R_{\delta
L}&=&\frac{1}{4}\big[2g_1\mathcal{H}\epsilon_{\alpha\beta\gamma}A_{\delta\beta}B_{\gamma
L}(AR)_{\alpha 4}-2g_2\mathcal{H}f_{IJK}B_{\delta J}C_{KL}(B^tR)_{I
4}\big]
\end{eqnarray}
where $B^3=\epsilon_{\alpha\beta\gamma}f_{IJK}B_{\alpha I}B_{\beta
J}B_{\gamma K}$. Before moving on, we note the useful relations
\begin{eqnarray}
R_{\alpha\beta}&=&\epsilon_{\alpha\beta\gamma}R_{\gamma4}, \qquad
(R^{\bar{K}(\bar{I}}R^{\bar{J})\bar{K}})_{a4}=3\delta^{\bar{I}\bar{J}}\delta_{a4},
\nonumber
\\
R^{\bar{K}(\bar{I}}_{i4}R^{\bar{J})\bar{K}}_{j4}&=&-\delta_{i\alpha}\delta_{j\alpha}\delta^{\bar{I}\bar{J}},
\qquad
R^{\bar{K}(\bar{I}}_{[i|l|}R^{\bar{J})\bar{K}}_{j]4}=\delta_{\alpha
l}\epsilon_{\alpha ij}\delta^{\bar{I}\bar{J}}, \nonumber
\\
R^{\bar{K}(\bar{I}}_{\alpha 4}R^{\bar{J})\bar{K}}_{\beta
4}&=&-\delta_{\alpha\beta}\delta^{\bar{I}\bar{J}}\, .
\end{eqnarray}
The following combination is useful in computing
$A_{2i}^{\bar{I}\bar{J}}$
\begin{eqnarray}
f^{\bar{K}(\bar{I}\phantom{a}j}_{\phantom{as}4,n+1}T^{\bar{J})\bar{K}}_j&=&\frac{3}{2}\delta^{\bar{I}\bar{J}}\Big(-g_1\mathcal{H}
\textrm{det}A+\frac{1}{6}g_2\mathcal{H}B^3+h_1\mathcal{H}^2\Big),\nonumber
\\
f^{\bar{K}(\bar{I}\phantom{a}j}_{\phantom{as}\delta
L}T^{\bar{J})\bar{K}}_j&=&
-\frac{3}{4}\delta^{\bar{I}\bar{J}}\mathcal{H}(g_1\epsilon_{\alpha\beta\gamma}\epsilon_{\delta\beta'\gamma'}A_{\beta\beta'}A_{\gamma\gamma'}B_{\alpha
L}\nonumber \\ & &-g_2f_{IJK}B_{\beta J}B_{\gamma
K}\epsilon_{\delta\beta\gamma}C_{IL}).
\end{eqnarray}
We then find $A_1$ and $A_2$ tensors
\begin{eqnarray}
A_1^{\bar{I}\bar{J}}&=&-2T\delta^{\bar{I}\bar{J}}, \nonumber \\
A_{2i}^{\bar{I}\bar{J}}&=&\frac{1}{2}T^{\bar{I}\bar{J}}_i+\frac{1}{6}X_i\delta^{\bar{I}\bar{J}}
\end{eqnarray}
where we have defined the following quantities
\begin{eqnarray}
T&=&2\Big(-g_1\mathcal{H}\textrm{det}A+\frac{1}{6}g_2\mathcal{H}B^3+\frac{1}{2}h_1\mathcal{H}^2\Big),\nonumber
\\X_{4,n+1}&=&\frac{3}{2}\Big(-g_1\mathcal{H}
\textrm{det}A+\frac{1}{6}g_2\mathcal{H}B^3+h_1\mathcal{H}^2\Big),\nonumber
\\
X_{\delta
L}&=&-\frac{3}{4}\mathcal{H}(g_1\epsilon_{\alpha\beta\gamma}\epsilon_{\delta\beta'\gamma'}A_{\beta\beta'}A_{\gamma\gamma'}B_{\alpha
L}\nonumber \\ & &-g_2f_{IJK}B_{\beta J}B_{\gamma
K}\epsilon_{\delta\beta\gamma}C_{IL}).
\end{eqnarray}
Using \eqref{CSV}, we can compute the potential
\begin{equation}
V=16T^2-2\Big(\frac{1}{4}T^{\bar{I}\bar{J}}_iT^{\bar{I}\bar{J}}_i+\frac{1}{9}X_iX_i\Big).
\end{equation}
After some manipulations, we can show that the resulting potential
is the same as \eqref{KKPotential} with the following
identifications
\begin{eqnarray}
\mathcal{H}\rightarrow
N^{\frac{1}{6}}e^{-4g}=e^{-\sqrt{2}\Phi},\qquad A \rightarrow \left(
  \begin{array}{c|c}
    \frac{1}{\sqrt{\mathbb{I}_{3}-\mathbf{A}^t\mathbf{A}}} & 0 \\ \hline
    0 & \cosh{\sqrt{2}\Phi} \\
  \end{array}
\right)
, \nonumber \\
B \rightarrow \left(
  \begin{array}{c|c}
    \frac{1}{\sqrt{\mathbb{I}_{3}-\mathbf{A}^t\mathbf{A}}}\mathbf{A}^t & 0 \\  \hline
    0 & \sinh{\sqrt{2}\Phi} \\
  \end{array}
\right), \qquad C \rightarrow \left(
  \begin{array}{c|c}
    \frac{1}{\sqrt{\mathbb{I}_{n}-\mathbf{A}\mathbf{A}^t}} & 0 \\  \hline
    0 & \cosh{\sqrt{2}\Phi} \\
  \end{array}
\right)\, .
\end{eqnarray}


\begin{thebibliography}{99}
\bibitem{nicolai1} H. Nicolai and H. Samtleben, ``Maximal gauged
supergravity in three dimensions'', Phys. Rev. Lett. \textbf{86}
(2001) 1686-1689, arXiv: hep-th/0010076.
\bibitem{nicolai2} H. Nicolai and H. Samtleben, ``Compact and noncompact gauged maximal
supergravities in three dimensions'', JHEP 0104 (2001) \textbf{022},
arXiv: hep-th/0103032.
\bibitem{nicolai3} T. Fischbacher, H. Nicolai and H. Samtleben,
``Non-semisimple and Complex Gaugings of $N=16$ Supergravity'',
Commun.Math.Phys. \textbf{249} (2004) 475-496, arXiv:
hep-th/0306276.
\bibitem{dewit} Bernard de Wit, Ivan Herger and Henning Samtleben, ``Gauged Locally Supersymmetric $D=3$ Nonlinear Sigma
Models'', Nucl. Phys. \textbf{B671} (2003) 175-216, arXiv:
hep-th/0307006.
\bibitem{csym} H. Nicolai and H. Samtleben, ``Chern-Simons vs Yang-Mills gaugings in three dimensions'', Nucl. Phys. B \textbf{638} (2002) 207-219
, arXiv: hep-th/0303213.
\bibitem{PopeSU2} H. L\"{u}, C. N. Pope and E. Sezgin, ``$SU(2)$ reduction of
six-dimensional (1,0) supergravity'', Nucl. Phys. \textbf{B668}
(2003) 237-257, arXiv: hep-th/0212323.
\bibitem{PopeSU22} H. L\"{u}, C. N. Pope and E. Sezgin,
``Yang-Mils-Chern-Simons Supergravity'', Class. Quant. Grav.
\textbf{21} (2004) 2733-2748, arXiv: hep-th/0305242.
\bibitem{SS} J. Sherk, J. H. Schwarz, ``How to get masses from extra
dimensions'', Nucl. Phys. \textbf{B153} (1979) 61.
\bibitem{ns2} H. Nicolai and H.Samtleben, ``Kaluza-Klein supergravity on $AdS_3\times
S^3$'', JHEP \textbf{09} (2003) 036. arXiv: hep-th/0306202.
\bibitem{gkn} Edi Gava, Parinya Karndumri and K. S. Narain, ``AdS$_3$ Vacua and RG Flows in Three Dimensional Gauged
Supergravities'', JHEP 04 (2010) \textbf{117}, arXiv: 1002.3760.
\bibitem{HS1} O. Hohm and H. Samtleben, ``Effective Actions for
Massive Kaluza-Klein States on $AdS_3 \times S^3 \times S^3$'', JHEP
\textbf{0505} (2005) 027. arXiv: hep-th/0503088.
\bibitem{nishino1} H. Nishino and E. Sezgin, ``The Complete $N=2$,
$D=6$ Supergravity With Matter and Yang-Mills Couplings'', Nucl.
Phys. \textbf{B144} (1984) 353.
\bibitem{nishino} H. Nishino and E. Sezgin, ``New Coupling of
Six-Dimensional Supergravity'', Nucl. Phys. \textbf{B505} (1997)
497, arXiv: hep-th/9703075.
\bibitem{Riccioni} F. Riccioni, ``All Couplings of Minimal Six-dimensional
Supergravity'', Nucl. Phys. \textbf{B605} (2001) 245-265.
\bibitem{Romans} L.J. Romans, ``Self-Duality for Interacting Fields: Covariant Field
Equations for Six Dimensional Chiral Supergravities'', Phys. Lett.
\textbf{B276} (1986) 71.
\bibitem{Sagnotti} A. Sagnotti, ``A Note on the Green-Schwarz Mechanism in Open String
theories'', Phys. Lett. \textbf{B294} (1992) 196, arXiv:
hep-th/9210127.
\bibitem{sen} A. Sen, ``M-theory on $(K_3\times S^1)/\mathbb{Z}_2$'', Phys. Rev. \textbf{53D} (1996)
6725, arXiv: hep-th/9602010.
\bibitem{green} M. B. Green, J. H. Schwarz and P. C.West, ``Anomaly-free chiral theories
in six dimensions'', Nucl. Phys. \textbf{B254} (1985) 327.
\bibitem{witten} M. J. Duff, R. Minasian, and E. Witten, ``Evidence for heterotic/heterotic duality'', Nucl.
Phys. \textbf{B465} (1996) 413-438, arXiv:hep-th/9601036.
\bibitem{duff} M. J. Duff and R. Minasian, ``Putting string/string duality to the test'',
Nucl. Phys. \textbf{B436} (1995) 507.
\bibitem{Pope3} H. L\"{u}, C. N. Pope and E. Sezgin,
``Group Reduction of Heterotic Supergravity'', Nucl. Phys.
\textbf{B772} (2007) 205-226, arXiv: hep-th/0612293.
\bibitem{new} Edi Gava, Parinya Karndumri and K. S. Narain, ``Two dimensional RG flows and Yang-Mills instantons'', in preparation.
\end{thebibliography}
\end{document}